# Risk Assessment of Transmission Lines Against Grid-ignited Wildfires

Saeed Nematshahi[1], Amin Khodaei[1], Ali Arabnya[1,2]
[1] Department of Electrical & Computer Engineering, Unversity of Denver, Denver, CO, USA
[2] Quanta Technology, Raleigh, NC 27607, USA
saeed.nematshahi@du.edu; amin.khodaei@du.edu; aarabnya@quanta-technology.com

*Abstract*—Wildfires ignited by the power lines have become increasingly common over the past decade. Enhancing the operational and financial resilience of power grids against wildfires involves a multifaceted approach. Key proactive measures include meticulous vegetation management, strategic grid hardening such as infrastructure undergrounding, preemptive de-energization, and disaster risk financing, among others. Each measure should be tailored to prioritize efforts in mitigating the consequences of wildfires. This paper proposes a transmission line risk assessment method for grid-ignited wildfires, identifying the transmission lines that could potentially lead to damage to the natural and built environment and to other transmission lines if igniting a wildfire. Grid, meteorological, and topological datasets are combined to enable a comprehensive analysis. Numerical analysis on the standard IEEE 30-bus system demonstrates the effectiveness of the proposed method.

*Keywords*—Fire ignition, risk assessment, grid resilience, system hardening, wildfire.

## I. INTRODUCTION

Wildfires are formidable natural phenomena with intense blazes that destroy diverse wildlands, including forests, grasslands, chaparral, savannahs, and shrublands. Wildfires can be ignited by both natural causes, such as lightning strikes, and anthropogenic activities, such as campfires and deliberate arson [1]. In recent years, the frequency and severity of wildfires have intensified, primarily due to two main factors. First, climate change and prolonged droughts make ecosystems more susceptible to the spread of fires. Second, population growth in a wildland-urban interface, as human settlements expand into natural habitats, raises the risk of fires igniting in wildlands, especially due to infrastructure development, notably the expansion of power lines [2].

Grid-ignited wildfires not only can lead to widespread power outages affecting a significant number of customers but also inflict irreversible damage upon the environment and civil structures. The 2024 Smokehouse Creek Fire was a record-breaking wildfire in northeastern Texas and western Oklahoma. It burned over 1 million acres and destroyed hundreds of structures, making it the largest wildfire in Texas's history and the largest in the US in 2024 [3]. The 2021 Dixie Fire in northern California burned around 1 million acres and made it the largest single source wildfire in California history. The 2018 Camp Fire was the deadliest and most destructive wildfire in California, with 85 deaths and 18,804 structures destroyed [4].

Electric utilities employ a range of preventive measures to mitigate the risk of grid-ignited wildfires. This includes regular vegetation management by trimming trees near power lines [5], conducting upgrades to ensure infrastructure resilience [6], and monitoring weather conditions to conduct preemptive power shutoffs in high-risk areas during fire weather conditions [7,8]. They also install fire detection systems along power lines for early detection [9], invest in community engagement initiatives to promote fire prevention and safety practices, and collaborate closely with emergency services for coordinated responses to wildfire events. These efforts collectively aim to reduce the likelihood of grid-ignited wildfires and protect both infrastructure and communities from the devastating impacts of such disasters.

Traditional wildfire risk assessments typically consider factors such as the likelihood and intensity of fires, with examples like the California Fire Hazard Severity Zone (FHZS) [10], the Wildfire Hazard Potential from the US Forest Service [11], andWiNGS which is a cloud-based tool that integrates visual representations of SDG&E's infrastructure with real-time weather data [12]. These assessments play a crucial role in evaluating relative risk and devising proactive strategies to mitigate wildfires, irrespective of their point of ignition. While effective for long-term planning, they often do not explicitly address ignition sources like fire-causing faults and failure modes in electric grid infrastructure, necessitating adaptation for comprehensive risk mitigation. There exists a significant research gap pertaining to the impact of grid-ignited wildfires and the necessity for conducting risk assessments to understand its implications. Power utilities, in particular, require a specific wildfire risk assessment to systematically consider the potential grid-ignited wildfire damage to both power infrastructure and the environment.

We aim to bridge this gap by leveraging landscape data, weather data, and power grid data to initiate desired ignition points on transmission lines and calculate the corresponding damages. Firstly, we assess the burned area to represent the environmental damage incurred. The burned area is the total land area that has been scorched by the wildfire. Secondly, we quantify the risk of each line cascading into other lines to fail, expressing the impact on the power grid. Finally, we integrate these findings into a novel metric that ranks transmission lines based on their associated risk. This novel metric can potentially provide electric utilities with valuable insights, enabling them to effectively plan for vegetation management, upgrading power infrastructure, undergrounding transmission lines, and optimizing their Public Safety Power Shutoff (PSPS) strategies.

The rest of the paper is organized as follows. Section II elaborates on the modeling of grid-ignited wildfires. In Section III, the numerical analysis using the IEEE 30-bus test system is presented, alongside a discussion of the proposed metric. Finally, Section IV concludes the paper and offers insights into future research directions.

## II. Grid-Ignited Wildfire Modeling

By considering a wildfire ignition point on a transmission line as well as the live fuel and weather data, the wildfire spread can be simulated, and accordingly, a high-fire threat district (HFTD) can be identified. By matching the HFTD area with the power grid topology, the impact of the wildfire on the grid will be obtained. Similarly, by matching the burned area with the civil structures, the impact of wildfire on the environment will be determined. This is shown in Figs. 1 and 2. Fig. 1 illustrates an arbitrary ignition point on a transmission line, shown with a red pin. Fig. 2 depicts the corresponding burned area due to this ignition, shown as a white area, alongside the lines being damaged due to the wildfire spread. Combining these two impacts that represent a potential risk, the overall consequence of the grid-ignited wildfire from each specific point on every transmission line could be measured as a wildfire risk metric. This risk metric can help rank the transmission lines in terms of hardening priority.

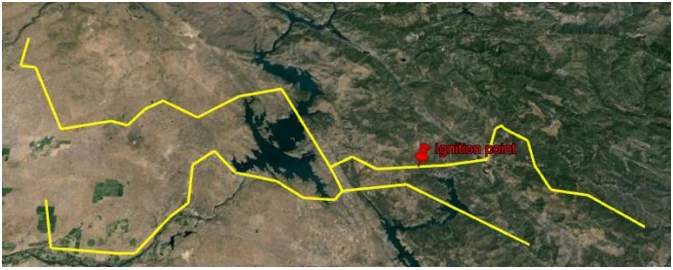

Fig. 1. Sample ignition point on a transmission line.

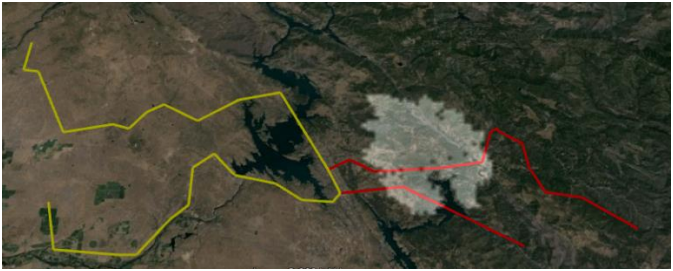

Fig. 2. Burned area and damaged power lines due to sample ignition.

The proposed method comprises the following four steps:

**Step 1:** Relevant datasets are collected, and ignition points are determined to create wildfire scenarios. The first set of data is landscape data including eight different factors: elevation, incline, direction of slope, fuel type, amount of tree cover, height of trees, lowest point of tree canopy, and density of tree canopy. The next set of data is weather data. This research incorporates wind direction, wind speed, temperature, and humidity, as they all play a significant role in wildfire propagation. Historical weather data for the study area is obtained from the National Solar Radiation Database (NSRDB) [13] at hourly intervals. The final piece of model input involves grid data. This encompasses a detailed topology of the transmission network, providing precise longitude and latitude coordinates for every element within the grid. At this stage, a number of ignition points on each line are selected, and then wildfire scenarios are sent as input to the wildfire simulator in Step 2.

**Step 2:** The created ignition points, along with collected data, are inputted into FARSITE software for wildfire spread simulation. FARSITE is a fire behavior simulator developed by the United States Forest Service. Primarily, FARSITE functions as a deterministic model, placing significant reliance on the data provided. Users furnish essential landscape details, encompassing terrain characteristics like slope and elevation, alongside fuel conditions such as vegetation type and moisture levels, as well as pertinent weather forecasts, including wind speed and humidity. Leveraging this comprehensive dataset, FARSITE employs established fire behavior models to forecast the propagation of fires, spanning ground-level spread to potential treetop engulfment, encompassing both surface and crown fire dynamics. The outcome of this step is the wildfire spread for each considered ignition point.

**Step 3:** The corresponding burned area for each scenario is matched with the landscape data, and the total acreage of the burned area is expressed. Similarly, the burned area is matched with the grid topology, and affected lines are obtained. Then, the average cost of a wildfire to the environment, encompassing property loss, timber damage, health-related costs, evacuation expenses, suppression efforts, and the average cost of transmission line reconstruction, are added to the program. At this stage, the model assesses the implications of environmental damage and additional damage to transmission lines.

The financial loss related to the burned environment as a result of an ignition by line $j$ is $LBE_j$ (Loss for Burned Environment) and calculated as follows:

$$LBE_j = \frac{\sum_{i=1}^{I} \sum_{c=1}^{C} S_c^i \cdot \propto \cdot CBE}{I} \quad (1)$$

where $j$, $i$, and $c$ represent the indices for the transmission line, ignition point, and cell, respectively. The size of each cell is contingent upon the desired resolution of the investigation. Denoted by $I$, the variable signifies the number of ignitions on each line, while $C$ represents the total number of cells within the study area. The cell status, labeled as $S$, assumes a value of 1 when the cell is burned and 0 when unaffected by the wildfire. The symbol $\propto$ denotes the ratio used to transform cell size to acreage, and $CBE$ stands for the average cost associated with the 1 acre of burned environment.

The financial loss related to the reconstruction of affected transmission lines, denoted by $LBL_j$ (Loss for Burned Lines) is:

$$LBL_j = \frac{\sum_{i=1}^{I} \sum_{j=1}^{J} l_j^i \cdot X_j \cdot CBL}{k} \quad (2)$$

where $J$ denotes the total number of lines, $l$ represents the status of affected lines, and $X_j$ denotes the length of line $j$. Additionally, $CBL$ represents the average cost of restoring the damage to a transmission line per mile.

**Step 4:** Finally, the model consolidates the financial losses to assess transmission line risk and introduces a novel metric to quantify the risk associated with each transmission line. The total wildfire financial loss ignited by an ignition point by line $j$, denoted by $WFL_j$, can be expressed as follows:

$$WFL_j = LBE_j + LBL_j \qquad (3)$$

The risk metric $M$ is expressed as the ratio of the total wildfire damage caused by an ignition by line $j$ to the worst-case scenario of a grid-ignited wildfire, and it is represented as:

$$M_j = \frac{WFL_j}{Max(WFL)} \qquad (4)$$

This metric seeks to evaluate the risk of a transmission line igniting a wildfire. The value of the metric ranges between 0 and 1, where a higher number indicates a higher potential risk for the line. Electric utilities, using this metric, can plan strategically to improve the resilience of their grid. Fig. 3 depicts the process, including the abovementioned steps.

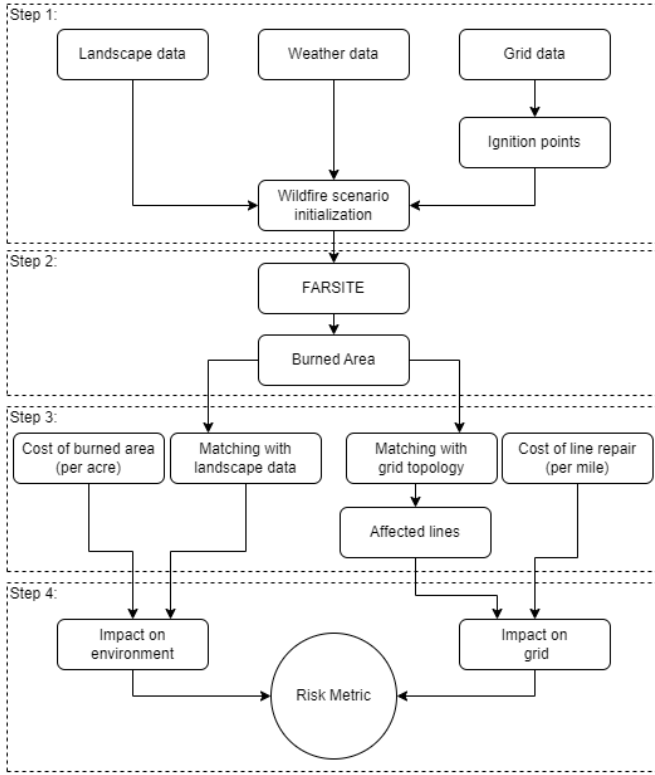

Fig. 3. Flowchart of the grid-ignited wildfire model.

## III. NUMERICAL ANALYSIS

We consider the IEEE 30-bus test system for a case study, which comprises 41 branches, consisting of 7 links and 34 lines, as depicted in Fig. 4. A geographic map of this system designed by the authors is overlayed with the landscape map and weather information. Fig. 5 provides a topographical representation of this system to enhance our understanding of the network—which is created based on the western area of Yosemite National Park and Stanislaus National Forest in California.

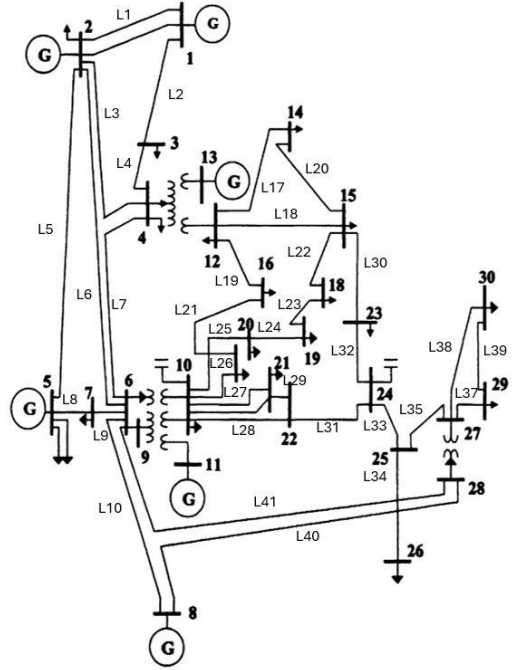

Fig. 4. Single line diagram of IEEE 30-bus test system.

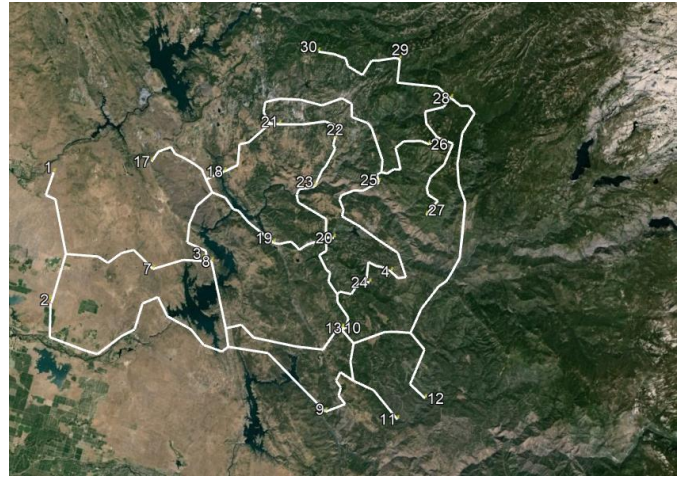

Fig. 5. Topology of the IEEE 30-bus Test System.

First, we evaluate the risk posed by each line to the surrounding environment in terms of potential fire damage. Next, we assess the risk of a wildfire ignition by a transmission line potentially affecting other lines. Finally, leveraging insights from both analyses, we numerically assess the metric aimed at quantifying the inherent risk associated with each transmission line in the event of a wildfire ignition.

We specifically collect weather information for all four seasons of 2022 separately to conduct a full-year risk analysis. Next, without loss of generality and for the sake of illustration, three ignition points are initiated on an *ad hoc* basis for each line to simulate a broad spectrum of potential scenarios, each representing a wildfire ignited by the respective line. The outcome for each line encompasses three distinct scenarios for each season, totaling twelve scenarios for each transmission line. Therefore, a grand total of 408 scenarios will be executed.

*A. Impact on environment*

Table I represents the average burned area for the three distinct wildfire scenarios corresponding to the ignition points on each line. As shown, each column represents the impact of weather conditions on the wildfire spread. The last column represents the average of these amounts for the four seasons of the year, representing the mean burned area.

As shown in Table I, line 6 is flagged as having the highest potential to ignite the largest wildfire, averaging a significant burned area of 4236.2 acres. The top five largest wildfires for each season are highlighted in bold. When considering the average values, lines 24, 8, 10, and 9 follow closely after line 6, representing the lines with the highest risk of damaging the environment. Additionally, it is evident that during winter, the areas of highest risk are delineated by lines 35, 37, 38, 33, and 34, setting it apart from other seasons. This divergence can be attributed to potential fluctuations in humidity, wind direction, and speed, which exert a distinct influence during this period.

It is worth noting that, on average, the highest burned area occurs during the summer months, while fall exhibits the next highest number of burns—a trend that aligns well with typical fire season patterns. This value for summer is 4.8 times higher than winter.

TABLE I. CORRESPONDING BURNED AREA (ACRES)

| Branch ID | Winter January 1st | Spring April 1st | Summer July 1st | Fall October 1st | AVG 2022 |
|---|---|---|---|---|---|
| 1 | 80.7 | 1267.0 | 3879.4 | 2163.9 | 1847.8 |
| 2 | 106.8 | 1497.2 | 4211.6 | 2360.9 | 2044.1 |
| 3 | 78.3 | 710.0 | 2299.2 | 1192.3 | 1069.9 |
| 4 | 163.7 | 1375.0 | 3460.6 | 2206.6 | 1801.5 |
| 5 | 281.2 | 1872.1 | 4637.5 | 2382.8 | 2293.4 |
| 6 | 322.1 | **3285.0** | **8230.4** | **5107.3** | **4236.2** |
| 7 | 346.4 | 1918.3 | **5392.0** | 3315.9 | 2743.2 |
| 8 | 1047.6 | **2440.3** | **6045.7** | **4371.7** | **3476.3** |
| 9 | 656.1 | 1837.7 | 4969.7 | **3620.8** | **2771.0** |
| 10 | 546.3 | **2114.1** | **6226.6** | **4347.4** | **3308.6** |
| 17 | 61.7 | 1066.5 | 3293.3 | 1800.9 | 1555.6 |
| 18 | 104.4 | 1575.5 | 3910.2 | 2531.7 | 2030.5 |
| 19 | 295.4 | 1195.8 | 3119.5 | 2153.2 | 1691.0 |
| 20 | 80.7 | 873.2 | 2680.0 | 1586.2 | 1305.0 |
| 21 | 688.7 | **1991.3** | 4354.5 | 3217.4 | 2563.0 |
| 22 | 201.7 | 1265.8 | 2872.2 | 2057.1 | 1599.2 |
| 23 | 98.5 | 321.5 | 1111.6 | 726.1 | 564.4 |
| 24 | 816.2 | **2926.7** | **7117.0** | **5164.2** | **4006.0** |
| 25 | 264.6 | 471.0 | 1440.2 | 952.6 | 782.1 |
| 26 | 277.6 | 857.7 | 3503.3 | 2095.1 | 1683.4 |
| 27 | 880.3 | 1216.0 | 2798.6 | 1719.0 | 1653.5 |
| 28 | 665.5 | 701.1 | 2195.9 | 1389.8 | 1238.1 |
| 29 | 708.3 | 498.3 | 843.5 | 589.0 | 659.8 |
| 30 | 237.3 | 746.2 | 1767.7 | 1398.7 | 1037.5 |
| 31 | 119.8 | 1053.5 | 3097.6 | 2159.2 | 1607.5 |
| 32 | 944.9 | 1352.4 | 4094.7 | 2817.0 | 2302.3 |
| 33 | **1987.1** | 1715.5 | 2672.9 | 1933.8 | 2077.3 |
| 34 | **1492.4** | 1245.7 | 1479.4 | 716.6 | 1233.5 |
| 35 | **2636.1** | 1842.4 | 1564.8 | 1016.7 | 1765.0 |
| 37 | **2392.9** | 1641.9 | 1388.0 | 805.5 | 1557.1 |
| 38 | **2341.9** | 1510.8 | 1332.9 | 788.3 | 1493.5 |
| 39 | 614.5 | 333.4 | 469.8 | 304.9 | 430.6 |
| 40 | 832.8 | 1110.4 | 3584.0 | 2056.0 | 1895.8 |
| 41 | 997.7 | 1150.8 | 2369.2 | 1354.8 | 1468.1 |

*B. Impact on other lines*

Table II presents data detailing the extent of transmission line reconstruction required following wildfires. As shown, a compilation of failed lines is generated by analyzing the intersection of the burned area with the transmission system's topology, and the total length of these affected lines is computed. This process is repeated for fire weather data across four seasons, and an average value is derived accordingly. The top five high-risk lines for each season are highlighted in bold.

At first glance, it may appear that the most damage to the grid would occur when the ignition point is on the longest line. However, line 40, which is the longest, does not rank among the top five in the right column of Table II. Because of its close proximity to other lines, a single line is more likely to cause significant damage to adjacent lines as well as itself, potentially leading to their failure and necessitating reconstruction. The highest damage is attributed to line 3, which results in 259.9 miles of damage. It is a long line and in the vicinity of other lines, which contributes to the high damage.

TABLE II. DAMAGE TO GRID (MILES)

| Branch # | Winter January 1st | Spring April 1st | Summer July 1st | Fall October 1st | AVG 2022 |
|---|---|---|---|---|---|
| 1 | 81.77 | 81.77 | 81.77 | 81.77 | 81.77 |
| 2 | 88.47 | 88.47 | 98.23 | 88.47 | 90.91 |
| 3 | **259.90** | **259.90** | **259.90** | **259.90** | **259.90** |
| 4 | 19.80 | 58.33 | 93.97 | 93.97 | 66.52 |
| 5 | **171.50** | **171.50** | **225.75** | **225.75** | **198.63** |
| 6 | **184.20** | **225.75** | **225.75** | **225.75** | **215.36** |
| 7 | 107.97 | 107.97 | 135.67 | 107.97 | 114.89 |
| 8 | 79.73 | 79.73 | 160.67 | 93.47 | 103.40 |
| 9 | 100.90 | **161.63** | **161.63** | **161.63** | **146.45** |
| 10 | 103.40 | 103.40 | **194.50** | 140.25 | **135.39** |
| 17 | 27.30 | 36.80 | 43.63 | 43.63 | 37.84 |
| 18 | 39.60 | 39.60 | 59.23 | 39.60 | 44.51 |
| 19 | 24.65 | 24.65 | 24.65 | 24.65 | 24.65 |
| 20 | 34.03 | 39.40 | 52.43 | 48.30 | 43.54 |
| 21 | 11.50 | 17.70 | 47.55 | 24.05 | 25.20 |
| 22 | 26.70 | 26.70 | 39.07 | 26.70 | 29.79 |
| 23 | 5.47 | 8.20 | 8.20 | 8.20 | 7.52 |
| 24 | 8.63 | 8.63 | 19.97 | 17.23 | 13.62 |
| 25 | 48.70 | 48.70 | 48.70 | 48.70 | 48.70 |
| 26 | 64.37 | 64.37 | 134.63 | 134.63 | 99.50 |
| 27 | 53.80 | 119.40 | 119.90 | 119.40 | 103.13 |
| 28 | 43.85 | 43.85 | 142.25 | 142.25 | 93.05 |
| 29 | 13.20 | 13.20 | 13.20 | 13.20 | 13.20 |
| 30 | 13.20 | 13.20 | 17.50 | 17.50 | 15.35 |
| 31 | 8.73 | 13.10 | 13.10 | 13.10 | 12.01 |
| 32 | 28.40 | 28.40 | 34.95 | 34.95 | 31.68 |
| 33 | 39.00 | 43.37 | 32.67 | 32.67 | 36.93 |
| 34 | 37.17 | 37.17 | 29.63 | 29.63 | 33.40 |
| 35 | 95.27 | 95.27 | 95.27 | 20.70 | 76.63 |
| 37 | **140.53** | 98.33 | 97.10 | 97.10 | 108.27 |
| 38 | 83.40 | 83.40 | 83.40 | 83.40 | 83.40 |
| 39 | 69.70 | 69.70 | 69.70 | 69.70 | 69.70 |
| 40 | **133.37** | **123.90** | 123.90 | 123.90 | 126.27 |
| 41 | 109.57 | 109.57 | 100.10 | 100.10 | 104.83 |

## C. Risk metric

The risk metric quantifies the monetary value of damages to both the environment and the power grid itself. We employed the total wildfire cost estimation per acre from [14] to convert the values in Table I, named damage to the environment, into financial losses. Given that the region under study falls within the range of estimates provided in the referenced study, without loss of generality, we assumed a total wildfire cost of $20,000 per burned acre for our rural study area. It is important to note that this number may vary depending on the destroyed structures, fatalities, and critical infrastructure within the wildfire zone. For example, if a wildfire spreads to an urban area, the costs will increase drastically. In either case, the proposed models are generic and can be used for any cost assumption. To account for damage to other transmission lines requiring reconstruction, we allocated $200,000 per mile for replacement costs.

The suggested risk metric, which quantifies the potential financial loss stemming from a wildfire ignited by each transmission line, highlights the high-risk lines that necessitate operational resilience strategies. Fig. 6 presents a comparative analysis of transmission lines using this metric.

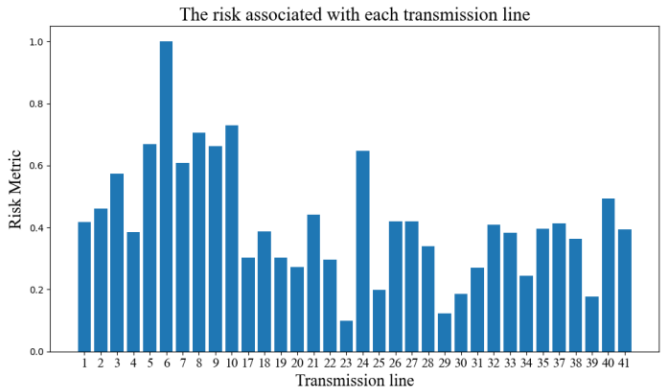

Fig. 6. Quantifying Transmission Line Risk using the proposed metric.

The analysis clearly identifies line 6 as the most critical candidate for immediate implementation of resilience strategies, given its highest risk metric of 1. As shown, line 10 emerges as the second most risky line, necessitating further risk reduction measures. Notably, lines 5, 6, 7, 8, 9, and 10 are situated in close proximity to each other, indicating a high concentration of risk for the electric utility. This clustering underscores the importance of prioritizing fire mitigation measures and grid hardening in this specific region to enhance overall resilience.

## IV. CONCLUSION AND FUTURE WORK

In this paper, we quantified the risk associated with a transmission line to ignite a wildfire. Topographical, meteorological, and power grid data are overlayed to assess the financial risk. Additionally, various weather data were considered in the model to provide comprehensive insights for grid operators. The proposed method not only investigates the risk of wildfires on the environment but also considers the risk to the power grid. Leveraging these factors, we proposed a novel metric to quantify the financial risk of wildfire ignition by each transmission line. The adoption of this wildfire risk metric can potentially help utilities with additional insights, empowering them to strategically plan for the allocation of their wildfire mitigation resources, including vegetation management, enhance power infrastructure, underground transmission lines, and optimization of their Public Safety Power Shutoff strategies with a data-driven, risk-informed approach. In future work, we will investigate the financial resilience of electric utilities against wildfire risk.